\documentclass[rnote]{aa} % for the research notes
\usepackage{graphicx}
\usepackage{txfonts}
\usepackage{natbib}
\bibpunct{(}{)}{;}{a}{}{,} % to follow the A&A style
\newcommand{\be}{\begin{equation}}
\newcommand{\ee}{\end{equation}}
\newcommand{\bea}{\begin{eqnarray}}
\newcommand{\eea}{\end{eqnarray}}

\begin{document}

  \title{ Mass-varying neutrino in light of cosmic microwave background and
    weak lensing}

  \author{G. La Vacca\inst{1}\and D. F. Mota\inst{2}}

  \institute{Physics Department ``G.~Occhialini'', Milano-Bicocca
    University \& I.N.F.N.,
% Sezione di Milano-Bicocca Piazza della  Scienza 3, 
20126 Milano, Italy
%\\ 
  %  \email{lavacca@mib.infn.it} 
    \and
    Institute of Theoretical Astrophysics, University of Oslo, 0315
    Oslo, Norway}
%\\ \email{d.f.mota@astro.uio.no} }

%   \date{Received September 15, 1996; accepted March 16, 1997}

% \abstract{}{}{}{}{} 
% 5 {} token are mandatory
 
  \abstract
      {}
      {We aim to constrain mass-varying neutrino models using
       large scale structure observations and produce forecast for the Euclid survey.}
      {We investigate two models with different scalar field potential and both positive and
       negative coupling parameters $\beta$. These parameters correspond to
       growing or decreasing neutrino mass, respectively. We explore
       couplings up to $|\beta|\lesssim 5$.}
      {In the case of the exponential potential, we find an upper limit
       on $\Omega_\nu h^2<0.004$ at 2-$\sigma$ level. In the case of
       the inverse power law potential the null coupling can be
       excluded with more than 2-$\sigma$ significance; the limits on
       the coupling are $\beta>3$ for the growing neutrino mass and
       $\beta<-1.5$ for the decreasing mass case. This is a clear sign
       for a preference of higher couplings. When including a prior on
       the present neutrino mass the upper limit on the coupling becomes
       $|\beta|<3$ at 2-$\sigma$ level for the exponential potential.
       Finally, we present a Fisher forecast using the tomographic weak
       lensing from an Euclid-like experiment and we also consider the combination
       with the cosmic microwave background (CMB) temperature and polarisation spectra from a
       Planck-like mission. If considered alone, lensing data is more
       efficient in constraining $\Omega_\nu$ with respect to CMB data
       alone. There is, however, a strong degeneracy in the
       $\beta$-$\Omega_\nu h^2$ plane. When the two data sets are
       combined, the latter degeneracy remains, but the errors are
       reduced by a factor $\sim2$ for both parameters.}
      {}

%   \keywords{cosmology --
 %    cosmological parameters --
  %   dark energy
   %}

   \titlerunning{MaVaN's in light of CMB and Weak Lensing
    } 

  \authorrunning{La Vacca $\&$ Mota}

   \maketitle
%
%________________________________________________________________
\section{Introduction}

%Present days astronomical observations strongly indicate there are two
%unknown components in the universe matter-energy budget. These are
%dark matter (DM) and dark energy (DE) \citep{Riess:1998cb,
%Perlmutter:1998np, Kowalski:2008ez}. The cosmological constant
%$\Lambda$ is one of the simplest candidates for the nature of dark
%energy. Although so far consistent with all major observational probes
%\citep{WoodVasey:2007jb, Reid:2009xm, Komatsu:2010fb,
%Sullivan:2011kv}, it faces two major puzzles: $1)$ Why such a tiny
%value (the cosmological constant problem)? $2)$ Why $\Lambda$ has
%become important only recently (the coincidence problem).  Several
%possible solutions to these issues have been put forward. In
%particular, by identifying dark energy with a scalar field
%\citep{Wetterich_1988, Ratra_Peebles_1988} and to couple it to either
%dark matter, or baryons or neutrinos
%\citep{Wetterich:1994bg,Amendola:1999er,Amendola:2011ie,
%LaVacca:2009yp,Kristiansen:2009yx}.

We study models in which the dark energy (DE) field is coupled
to neutrinos  \citep{Brookfield1, Brookfield2,
Brookfield3, Amendola_Baldi_Wetterich_2008, grow3}.
%grow4, grow6,
%grow7, grow8, grow9, grow10, grow11}.
%\citep{mavansde1, mavansde2, mavansde3, mavansde4,
%mavansde5, mavansde6, mavans1, mavans2, mavans3, mavans4, mavans5}.
%
%\cite{Amendola07} have showed that the coincidence problem can indeed
%be solved for a mass growing neutrino with a coupling to quintessence
%somewhat larger than gravity.
Several authors have shown that important cosmological effects can
appear within this class of models, especially when the neutrino
mass is sufficiently big for neutrinos to be non-relativistic\citep{grow5,grow2,grow5,Pettorino:2010bv}. 
%
% When
%this happens, neutrinos feel the presence of the fifth force and can
%collapse into nonlinear structures, which are stable and bounded
% It has been shown that these
%neutrino lumps form at redshift $z_\text{nl} \approx 1-2$, when the
%neutrino fluctuations become nonlinear \citep{grow1,grow2,
%grow5}. These effects are mainly present in the strong coupling regime
%where a simple linear approach may be insufficient
%\citep{}. Here we limit our analysis to the low
%coupling regime in which such instabilities are absent.  
In this work, we further investigate the cosmological signatures of the
mass-varying neutrino (MaVaN) models and put constraints on them using
the most updated observational data from the cosmic microwave
background (CMB) radiation temperature anisotropies spectra, from the
large scale structure (LSS), and the supernovae type Ia (SNIa)
luminosity distance. We start by investigating two different scalar
field potentials and both positive and negative coupling
parameters. These parameters corresponds to growing or decreasing
neutrino mass, respectively. Finally, we present a Fisher forecast
using the tomographic weak lensing from an Euclid-like experiment
\citep{Laureijs:2011mu} in the last section. This is also shown in
combination with the CMB temperature and polarisation spectra from a
Planck-like mission \citep{Planck:2006aa}.

\section{The cosmological background evolution}
%\section{Theory}
%\subsection{The Cosmological Background Evolution}

The MaVaN models involve a coupling between the DE scalar field and
massive neutrinos. 
%In a flat, homogeneous, Friedmann-Robertson-Walker
%universe with line-element
%\begin{equation}
%ds^2 = a^2(\tau)\left(-d\tau^2 + \delta_{ij}dx^i dx^j \right),
%\end{equation}
%the Friedmann equation describes the universe expansion: \be
%{\cal H}^2\equiv\left(\frac{a'}{a}\right)^2 = \frac{a^2}{3}
%\sum_\alpha \rho_\alpha~. \ee In this equation, $\rho_\alpha(\tau)$ is
%the energy density of the individual components $\alpha$, including
%CDM, DE, neutrinos, baryons and radiation. The prime refers to the
%derivative with respect to conformal time $\tau$.
%
The coupling of DE to the neutrinos results in the neutrino mass
becoming a function of the scalar field,
%, whose evolution is described
%by the coupling expression \be
%\label{beta_phi} \beta \equiv - \frac{d \ln{m_\nu}}{d \phi}~.
%\ee 
%We choose to take a constant $\beta$ such that:
\be 
\label{eq:nu_mass} m_{\nu} \equiv \bar{m}_{\nu} e^{{{\beta}} \phi}~,
\ee where $\bar{m}_{\nu}$ is a constant. An additional attractive
force between neutrinos of strength $2\beta^2$ is mediated by the
scalar field exchange. 
%For $\beta \sim 1$ this corresponds to a
%strength comparable to gravity.
%
The 
%existence of the coupling between DE and massive neutrinos can be
%formalised, assuming that the sum of the energy momentum tensors for
%the two species is conserved, but not the separate parts. We neglect a
%possible scalar field coupling to CDM, so that $\label{cons_cdm}
%\rho_c' = -3 {\cal H} \rho_c $. This hypothesis requires 
DE and mass varying neutrinos obey the coupled conservation equations: \bea
\label{cons_phi} \rho_{\phi}' = -3 {\cal H} (1 + w_\phi) \rho_{\phi} +
\beta \phi' (1-3 w_{\nu}) \rho_{\nu} \\
\label{cons_gr} \rho_{\nu}' = -3 {\cal H} (1 + w_{\nu}) \rho_{\nu} - 
\beta \phi' (1-3 w_{\nu}) \rho_{\nu} \eea where the energy density
and pressure stored in the neutrinos is given by \be\label{eq:density}
\rho_\nu=\frac{1}{a^4}\int q^2 dq\, d\Omega \epsilon f_0(q), \quad p_\nu=\frac{1}{3a^4}\int q^2
dq\, d\Omega f_0(q) \frac{q^2}{\epsilon}. \ee Here $f_0(q)$ is the
usual unperturbed background neutrino Fermi-Dirac distribution
function: \be f_0(\epsilon)=\frac{g_s}{h_P^3}\frac{1}{e^{\epsilon/k_B
T_0}+1},\ee and $\epsilon^2 = q^2 + m_\nu^2(\phi)a^2$ ($q$ denotes the
comoving momentum). As usual, $g_s$, $h_P$, and $k_B$ stand for the
number of spin degrees of freedom, Planck's constant, and Boltzmann's
constant, respectively. 
%In the following we will assume that the
%neutrinos decouple whilst they are still relativistic, and therefore
%the phase-space density only depends upon the comoving momentum.

%The energy density and pressure of the quintessence field are given by  \be
%\label{phi_bkg} \rho_{\phi} = \frac{\phi'^2}{2 a^2} + V(\phi), \,\,\,
%p_{\phi} = \frac{\phi'^2}{2 a^2} - V(\phi), \,\,\, w_{\phi} =
%\frac{p_{\phi}}{\rho_{\phi}}~. \ee 
Taking the energy conservation of the coupled neutrino-dark energy system into 
account, one can immediately find that the evolution of the scalar field is described
by a modified Klein-Gordon equation: \be \label{kg} \phi'' + 2{\cal H}
\phi' + a^2 \frac{dV}{d \phi} = a^2 \beta (\rho_{\nu}-3 p_{\nu})~. \ee
%This equation contains an extra source term with respect to the
%uncoupled case, which accounts for the energy exchange between the
%neutrinos and the scalar field.

Here we investigate two different expressions for the DE potentials, an
exponential potential \citep{Wetterich_1988} and an inverse power-law \citep{Binetruy:1998rz}: \be
\label{pot_def} V(\phi) \propto e^{- \sigma \phi}, \qquad V(\phi)
\propto (M/\phi)^2~.\ee 
%Both $\sigma$ and $M$ are constant. 
%In
%particular $\sigma$ is a free parameter which determines the slope of
%the potential and thus the DE fraction at early times. On the other
%hand, $M$ defines the potential energy scale and, together with
%$\bar{m}_{\nu}$ in Eq.~\ref{eq:nu_mass}, is used to set the
%convergence of the model to the actual cosmology.

%Once the potential (\ref{pot_def}) or (\ref{powpot}) is given, the
%evolution equations can be numerically solved. 
%HERE 
The initial conditions for the field $\phi$ and its time
derivative are chosen in a recursive way to reproduce the
correct value of the assigned density parameters at $a=1$.  In the
very early universe, neutrinos are still relativistic and almost
massless, where $p_\nu = \rho_\nu/3$ and the coupling term in
Eqs.(\ref{cons_phi}), (\ref{cons_gr}), (\ref{kg}) vanishes.
%
%%%%%%%%%%%%%%%%%%%%%%%%%%%%%%%%%%%%%%%%%%%%%%%%%%%%%%%%%%%%%%%%%%%%
%\begin{figure}%[!h]
%\begin{center}
%\includegraphics[height=6.cm,width=8.cm,angle=0]{densities.eps}
%\end{center}
%\caption{Energy density evolution of different components, as
%specified in the legend, for the potential (\ref{pot_def}), with
%$\beta=-1.9$, $\sigma=0.021$, $\omega_\nu=0.0010$,
%$\omega_{CDM}=0.1097$, $\omega_b=0.0224$, $h_0=70$.}
%\label{fig_1_const}
%\end{figure}
%%%%%%%%%%%%%%%%%%%%%%%%%%%%%%%%%%%%%%%%%%%%%%%%%%%%%%%%%%%%%%%%%%%%
%
However, the coupling term $\sim \beta \rho_\nu$ becomes significantly 
different from zero as soon as neutrinos become nonrelativistic,
affecting the evolution of the field $\phi$. 
%As shown in Fig.~{\ref{fig_1_const}}, $\rho_\nu$ and $\rho_\phi$
Neutrinos and $\phi$ densities change behaviour for $z < 6$: the value
of the scalar field stays almost constant, and the frozen scalar
field potential mimics a cosmological constant. According to this
model, neutrinos at the present time are still subdominant with
respect to cold dark matter (CDM), though they will take the lead in 
the future. For our choice of the parameters, neutrino pressure terms may 
be safely neglected for redshifts $z_{nr}<4$; before that, redshift neutrinos
free stream as usual relativistic particles.

\subsection{Perturbed equations}

%In order to compute the evolution of cosmological
%perturbations in
%our model. we work in the synchronous gauge, taking the line element
%to be \be ds^2 = -a^2 d\tau^2 +a^2 \left(\delta_{ij} +h_{ij}\right)
%dx^i dx^j.\ee 
The perturbed Klein-Gordon equation is given by
\citep{Brookfield2}: 
\begin{eqnarray}\label{pertkg}
\lefteqn{\ddot{\delta \phi}+ 2H \dot{\delta \phi}+\left(k^{2} +
a^{2}\frac{d^{2}V}{d\phi^{2}}\right)\delta \phi+\frac{1}{2}\dot{h}
\dot{\phi}=} \\ \nonumber & &-a^2 \left[\frac{d\ln
m_\nu}{d\phi}(\delta\rho_\nu-3\delta p_\nu)+\frac{d^{2} \ln
m_\nu}{d\phi^{2}}\delta\phi(\rho_\nu-3 p_\nu)\right].
\end{eqnarray} 

For the neutrinos, we use the perturbed component of the energy momentum
conservation equation for the coupled neutrinos 
%\be
%T^\mu_{~~\gamma;\mu} =\frac{d \ln m_\nu}{d\phi} \phi_{,\gamma}
%T^{\alpha}_{~\alpha} \ee 
%to calculate the evolution equations for the
%neutrino perturbations ($T^{\alpha}_{~\alpha}$ stands for the trace of
%the neutrino energy momentum tensor). Taking $\gamma=0$ we derive the
%equation governing the evolution of the neutrino density contrast ,
$\delta_\nu \equiv \frac{\delta \rho_\nu}{\rho_\nu}$, whilst taking
$\gamma=i$ (spatial index) yields the velocity perturbation equation
$\theta_\nu \equiv ik_{i}v^{i}_{\nu}$ with the coordinate velocity
$v^{i}_{\nu} \equiv dx^{i}/d {\tau}$: \bea\label{denscont}
\dot{\delta}_\nu &=& 3\left(H +\beta \dot{\phi}\right)\left (w_\nu-
\frac{\delta p_\nu}{\delta \rho_\nu}\right)\delta_\nu - \left(1
+w_\nu\right)\left (\theta_\nu + \frac{\dot{h}}{2}\right)\nonumber \\
&+& \beta\left(1-3w_\nu\right) \dot{\delta \phi} +\frac{d\beta}{d\phi}
\dot{\phi} \delta\phi \left(1-3w_\nu\right),\eea \bea \dot{\theta_\nu}
&=& -H(1-3w_\nu)\theta_\nu -\frac{\dot{w_\nu}}{1+w_\nu} \theta_\nu
+\frac{\delta p_\nu / \delta \rho_\nu}{1+w_\nu}k^{2}\delta_\nu
\nonumber \\ &+& \beta\frac{1-3w_\nu}{1+w_\nu} k^{2} \delta\phi
-\beta(1-3w_\nu) \dot{\phi}\theta_\nu - k^{2} \sigma_\nu. \eea 
%The
%variable $\sigma_\nu$ represents the neutrino anisotropic stress. It
%is the presence of the additional coupling terms in these expressions
%for the growth of the neutrino density and velocity perturbations, as%
%well as the modifications to the evolution of the cosmological
%background, which alters the behaviour of the neutrino perturbations
%in comparison with the standard uncoupled case.

\section{Methods and data}

In this paper, we use two methods to test MaVaN theory against
cosmological data: the likelihood analysis through Markov-Chain
Monte-Carlo (MCMC) technique and the Fisher information matrix.

%The CMB temperature and polarization anisotropy spectrum and matter
%power spectrum are calculated by suitably modifying CAMB \citep{camb}
%to contemplate MaVaN's equations as described above. To ensure the
%accuracy of our calculations, we directly integrate the neutrino
%distribution function, rather than using the standard velocity
%weighted series approximation scheme.

The CMB anisotropy and matter power spectra are calculated by suitably
modifying CAMB \citep{camb} to include MaVaN's equations as
described above. To ensure the accuracy of our calculations, we
directly integrate the neutrino distribution function, rather than
using the standard velocity weighted series approximation scheme.

For the MCMC analysis, we use a modified version of the publicly
available codes {\sc CosmoMC} \citep{Lewis:2002ah} to explore the
parameter space. We consider the following basic set of parameters:
$$\{\omega_b, \omega_{CDM}, H_0, z_{re}, {\rm n_s}, \ln {\rm A_s},
\omega_\nu, \beta\}$$ and also $\sigma$ when the exponential potential
is involved. Here, $\omega_{b}$, $\omega_{CDM}$, $\omega_{\nu}$ are
the physical baryon, cold dark matter, and neutrino density
parameters. Here, $\omega_{i} = \Omega_{i} h^2$, where
$i=b,~CDM,~\nu$, and $h$ is the dimensionless Hubble parameter $H_0$;
$z_{re}$ is the redshift of reionization; $n_s$ is the scalar spectral
index; $A_s$ denotes the amplitude of the scalar fluctuations at a
scale of $k=0.05$~Mpc$^{-1}$. The sum of the $\nu$ masses is directly
related to the neutrino density parameter through the relation $M_\nu
= \Sigma m_\nu = \omega_\nu \cdot 93.5$ eV, assuming three equal mass
$\nu$ values.

All parameters are given flat priors, unless otherwise stated
explicitly. We separately consider positive and negative $\beta$
values in Eq.~(\ref{eq:nu_mass}), which correspond to either a growing
or decreasing neutrino mass, respectively.
%These are very different
%physical theories, from the point of view of both the high energy
%physics behind as well as the cosmology and astrophysics. 
We choose to
span only the low coupling regime up to $|\beta|<5$, since linear
perturbations become unstable for high couplings, {\it e.g.} $|\beta|
\sim 50$ \citep{grow2, grow5}.
%,mavansde4, Franca:2009xp}..
%
%This is not a new issue in MaVaNs models, where many other
%authors have reported those instabilities However, there are several other claims in
%the literature that as the neutrino perturbations become nonlinear,
%those structure virialise and become stable \citep{Pettorino:2010bv,
%baldi}. Hence, the strongly coupled mass-varying neutrino models may
%not be ruled out by the large scale structure formation in the linear
%regime, such as the integrated Sachs-Wolfe effect or the matter power
%spectra. In spite of the many attempts to understand and manage those
%``non-linearities'' \citep{Pettorino:2010bv,baldi}, a proper method of
%coping with these effects in a MCMC method to probe the parameter
%space of the models in the high $\beta$ regime is still
%lacking. Therefore, we leave the strong coupling regime for a future
%work. 
In our MCMC analysis, we assume that the Universe is spatially
flat.

With the aim of obtaining the best estimate of the cosmological
parameters, we combine different CMB data sets (WMAP
\citep{Komatsu:2010fb}, CBI \citep{Sievers:2005gj}, ACBAR
\citep{Reichardt:2008ay}, VSA \citep{Dickinson:2004yr}) with data from
LSS \citep{Tegmark:2006az} and SNIa \citep{Amanullah:2010vv}.  We also
apply additional priors on the Hubble parameter of $H_0 = 74.2 \pm
3.6$ \citep{Riess:2009pu}.

%With the aim of obtaining the best estimate of the cosmological
%parameters, we combine different CMB datasets with data from large
%scale structure and SNIa. In particular, for the CMB temperature
%anisotropy spectra we considered the 7 year data from the WMAP
%satellite (WMAP7, see {\it e.g.}  \cite{Komatsu:2010fb}), plus higher
%multipole data from CBI \citep{Sievers:2005gj}, ACBAR
%\citep{Reichardt:2008ay}, VSA \citep{Dickinson:2004yr}
%experiments. For the matter power spectrum we choose the results from
%the SDSS survey \citep{Tegmark:2006az}. Constraints on the recent
%expansion history of the Universe are given by the SNIa observations
%from the Union2 compilation \citep{Amanullah:2010vv}. We also apply
%additional priors on the Hubble parameter $H_0 = 74.2 \pm 3.6$
%\citep{Riess:2009pu}.

In addition to the likelihood analysis, we test the power of future
CMB and weak lensing data by constraining MaVaN's parameters using the
Fisher formalism \citep{Fisher:1935,Tegmark:1996bz}.

\section{MCMC results}

%\subsection{Cosmological data}
\label{par:4.1}

%In the case of the exponential potential (\ref{pot_def}), likelihood
%contours and distributions for growing ($\beta>0$) or decreasing
%($\beta<0$) $\nu$ mass are shown in Fig.~\ref{mav01}. We report plots
%for $\omega_\nu$, $\sigma$ and $\beta$ which are the most significant
%parameters in our models, the constraints on the other parameters not
%considerably differing from the $\Lambda$CDM case.

\begin{figure}%[!h]
\begin{center}
\includegraphics[height=5.cm,angle=0]{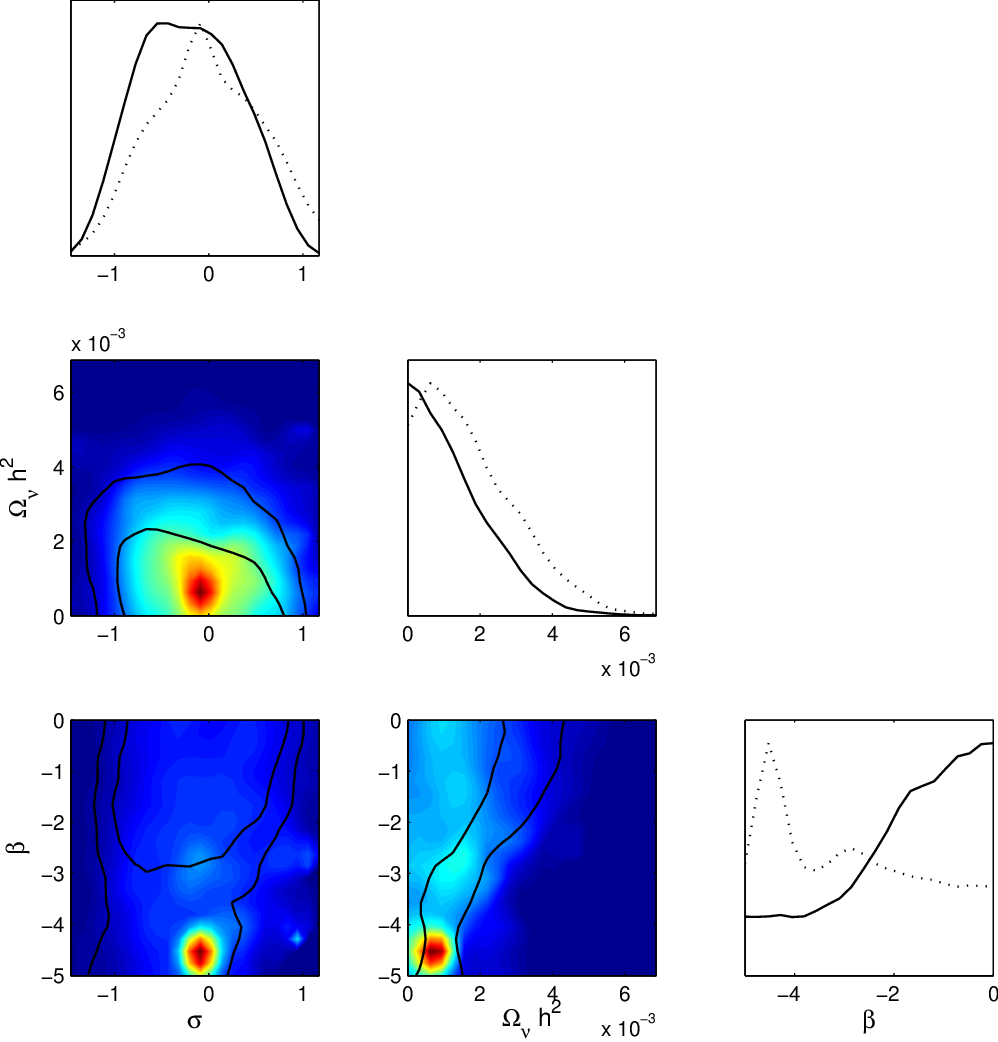}
\vskip.1truecm
\includegraphics[height=5.cm,angle=0]{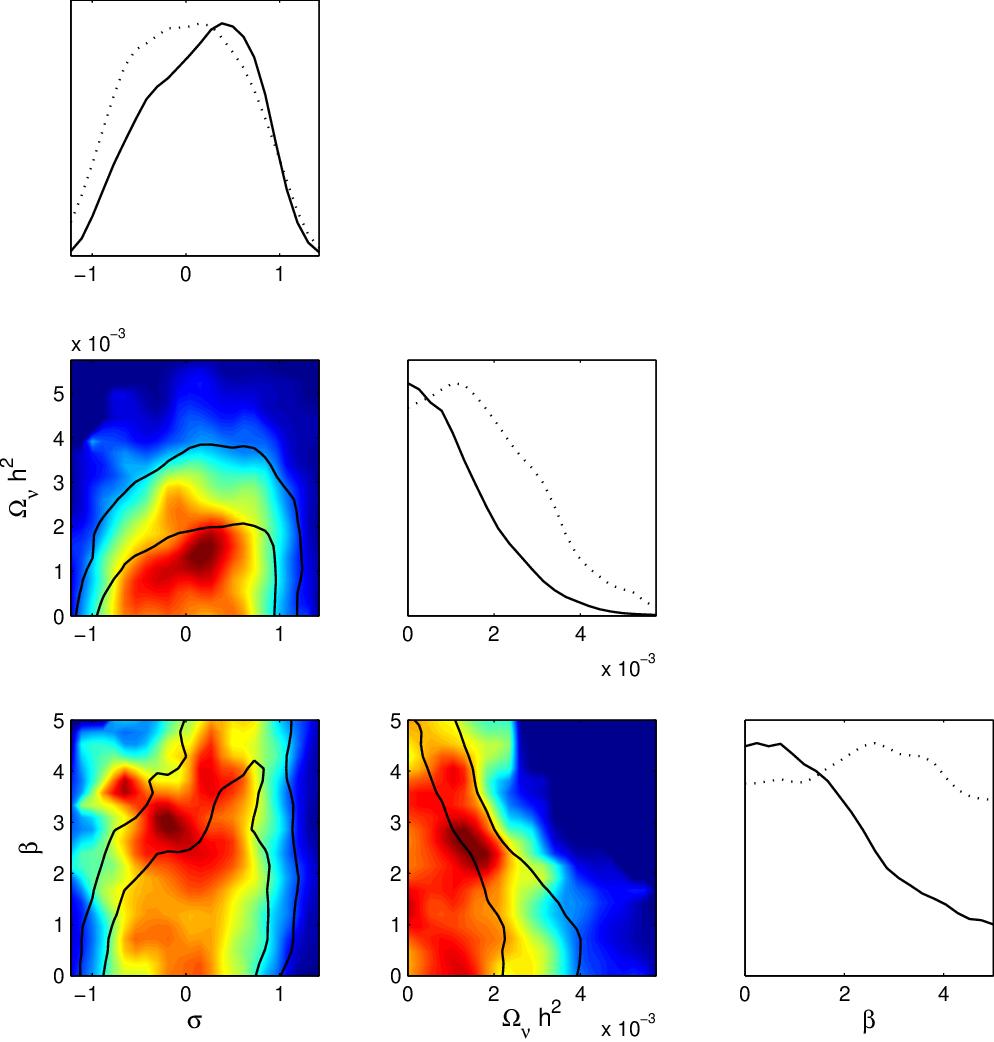}
\end{center}
\caption{In the upper (lower) panel, likelihood contours and
  distributions are shown for decreasing (growing) neutrino mass in
  the exponential potential case. For the plots on the diagonal,
  dotted (solid) lines are mean (marginalized) likelihoods of samples.
%, solid lines are marginalised probabilities. 
Similarly, black lines on the plots exhibit 1- and 2-$\sigma$
contours of the marginalised probability distribution, while the
colours refer to the mean likelihood degradation from the top (dark red)
to lower values.}
\label{mav01}
\end{figure}

In the case of the exponential potential (\ref{pot_def}), the plots
for $\omega_\nu$, $\sigma$, and $\beta$ are shown in
Fig.~\ref{mav01}. For both a growing and decreasing mass, no stringent
constraints can be put on the coupling $\beta$, whose upper limit
exceeds $|\beta|=5$ at 68\% confidence level (CL). In contrast, the
$\nu$ density parameter is well constrained at 95\% CL: $\omega_\nu <
0.0034$ ($\omega_\nu < 0.0032$) for a decreasing (growing) mass.
%$\omega_\nu < 0.0034$ (decreasing mass) and $\omega_\nu < 0.0032$
%(growing mass).
These limits are $\sim3$ times narrower than those found in
\cite{Brookfield2}. In terms of current neutrino mass these values result in
an upper limit to $m_\nu \simeq 0.32$eV. 
%The limits on $\sigma$ are
%centered around zero in both cases: $-1.12<\sigma<0.77$
%($-0.89<\sigma<1.05$) for decreasing (growing) mass.
%$-1.12<\sigma<0.77$ (decreasing mass) and $-0.89<\sigma<1.05$ (growing
%mass) at 1-$\sigma$.
We note here that the results show a common behaviour in the
$\beta$~{\it vs.}~$\omega_{\nu}$ plots, which are almost symmetric with respect
to the sign of $\beta$. A degeneracy appears between the two
parameters, so that a high $|\beta|$ agrees with low neutrino
content, or, equivalently, low neutrino mass.

%%%%%%%%%%%%%%%%%%%%%%%%%%%%%%%%%%%%%%%%%%%%%%%%%%%%%%%%%%%%%%%%%%%%
%%%%%%%%%%%%%%%%%%%%%%%%%%%%%%%%%%%%%%%%%%%%%%%%%%%%%%%%%%%%%%%%%%%%

\begin{figure}%[!h]
\begin{center}
\includegraphics[height=5.cm,angle=0]{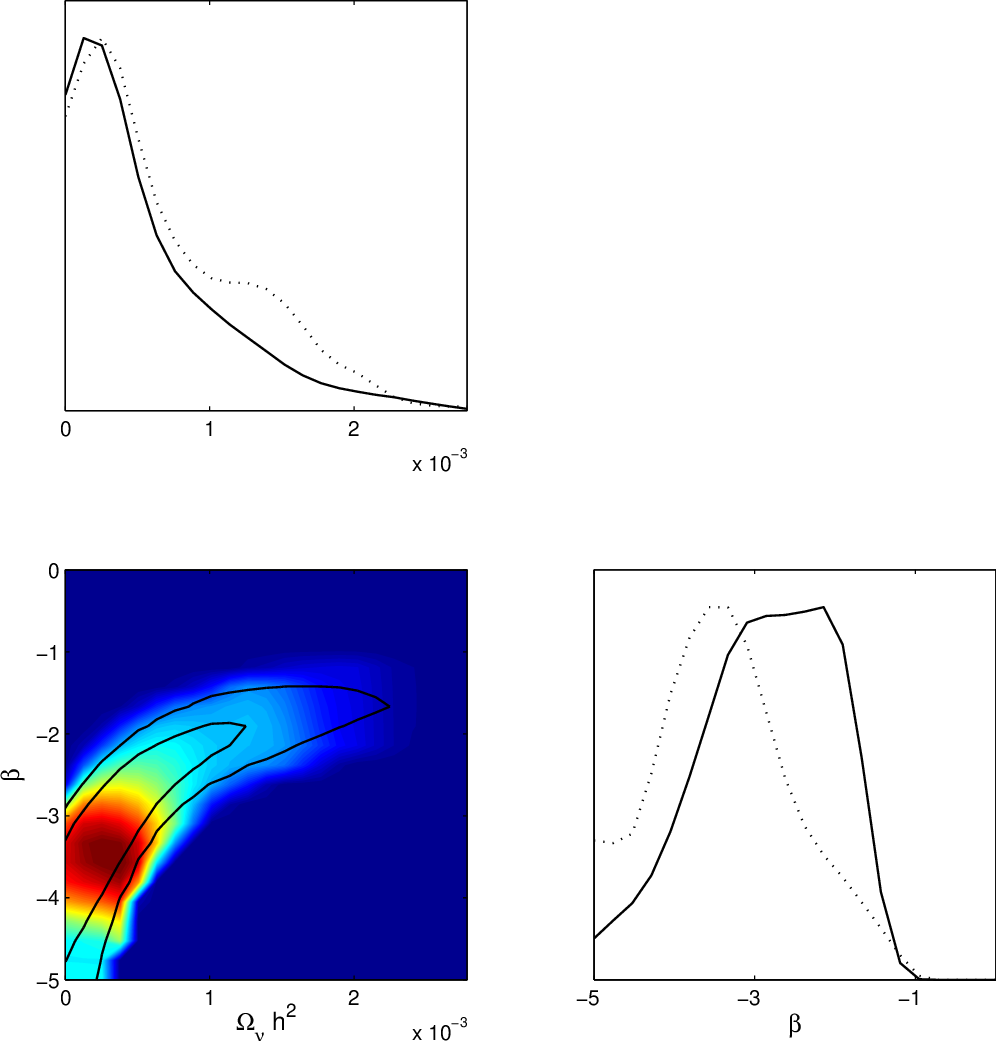}
\vskip.1truecm
\includegraphics[height=5.cm,angle=0]{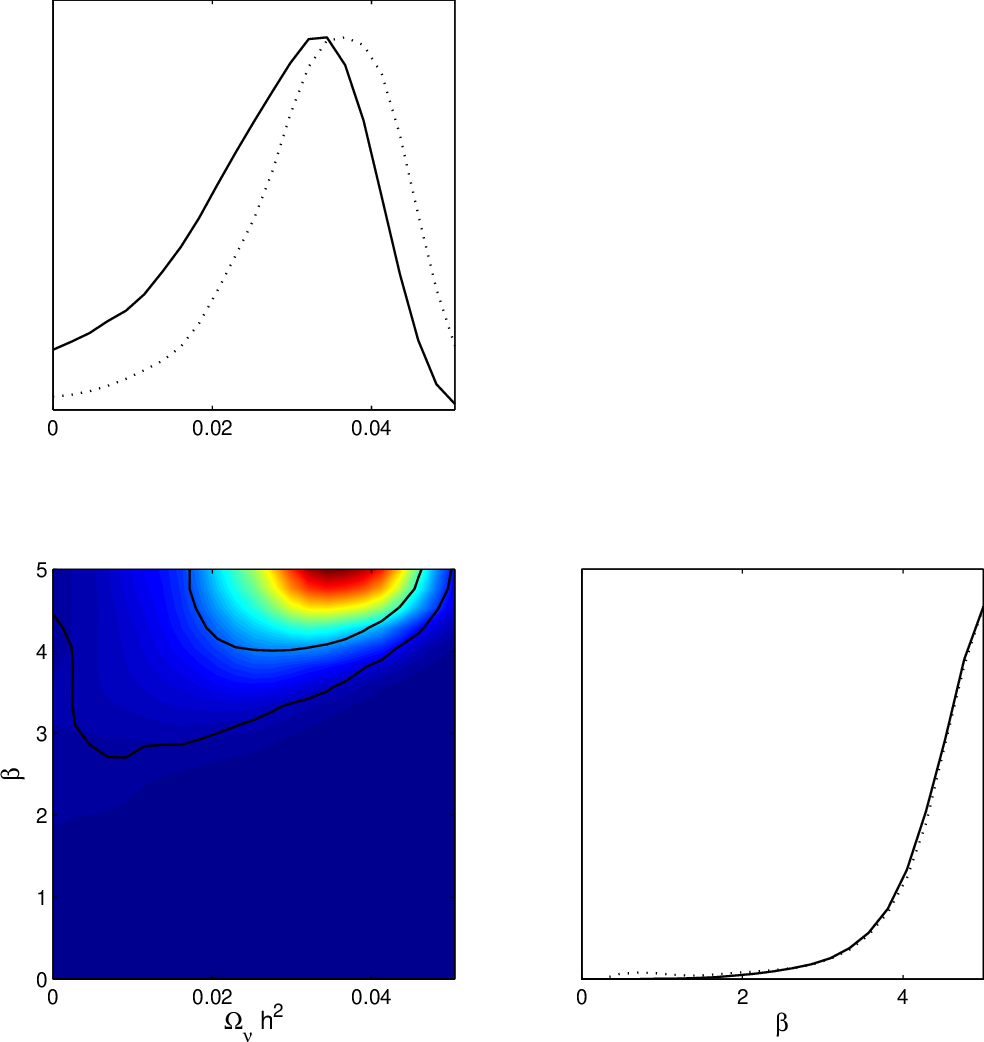}
\end{center}
\caption{As in Fig.~\ref{mav01}, when the power-law potential is
considered.}
\label{mav10}
\end{figure}

In Fig.~\ref{mav10}, we reported likelihood contours for the case of
the inverse power-law potential. This potential type was not
previously considered in \cite{Brookfield2}. In the $\beta$~{\it
  vs.}~$\omega_{\nu}$, plots the zero coupling is apparently excluded
with statistical significance that is higher than 95\% confidence level (CL),
with $\beta>3$ for the growing neutrino mass and $\beta<-1.5$ for the
decreasing mass case at 2-$\sigma$. In particular, the growing
neutrino mass case exhibits an explicit preference for high $\beta$
values, which agree with \cite{Amendola07}. For $\omega_{\nu}$, only
upper limits can be found at 95\% CL: $\omega_{\nu}<0.0017$
($\omega_{\nu}<0.0423$) for decreasing (growing) mass.
%$\omega_{\nu}<0.0017$ (decreasing mass) and $\omega_{\nu}<0.0423$
%(growing mass).

%%%%%%%%%%%%%%%%%%%%%%%%%%%%%%%%%%%%%%%%%%%%%%%%%%%%%%%%%%%%%%%%%%%%
%%%%%%%%%%%%%%%%%%%%%%%%%%%%%%%%%%%%%%%%%%%%%%%%%%%%%%%%%%%%%%%%%%%%

%\subsection{External priors}

%Complementary information to cosmological data can be obtained by
%taking into account priors on parameters coming from external
%measurements. In such a way a better knowledge of the other parameters
%can be obtained.

To exploit complementary information from external
measurements, we performed a specific analysis,
%
%In our analysis we fix 
by fixing the current neutrino mass value
%(and as a consequence $\omega_\nu$), 
using two possible options, $m_\nu=0.2$~eV and $m_\nu=0.3$~eV~. These
values are within the range of the claimed $\nu_e$ mass detection in
the Heidelberg-Moscow experiment
\citep{KlapdorKleingrothaus:2004wj,KlapdorKleingrothaus:2005qv}. The KATRIN 
experiment \citep{Sturm:2011ms} is also expected to constrain
the value of the neutrino mass with a sensitivity in the sub-eV range.
Besides fixing $\omega_\nu$, we chose to fix 
%the baryon density parameter 
$\Omega_b h^2 = 0.022$, as constraints on this parameter are not
significantly modified by neutrino-DE couplings, and %. Also the value for
$H_0 = 72$~km~s$^{-1}$Mpc$^{-1}$.% is kept constant.

%In the exponential potential case we are therefore left with a
%cosmological model requiring six parameters: $\beta$, $\sigma$,
%$\Omega_{CDM} h^2$, $z_{re}$, $A_s$ and $n_s$.

\begin{figure}%[!h]
\begin{center}
\includegraphics[height=5.cm,angle=0]{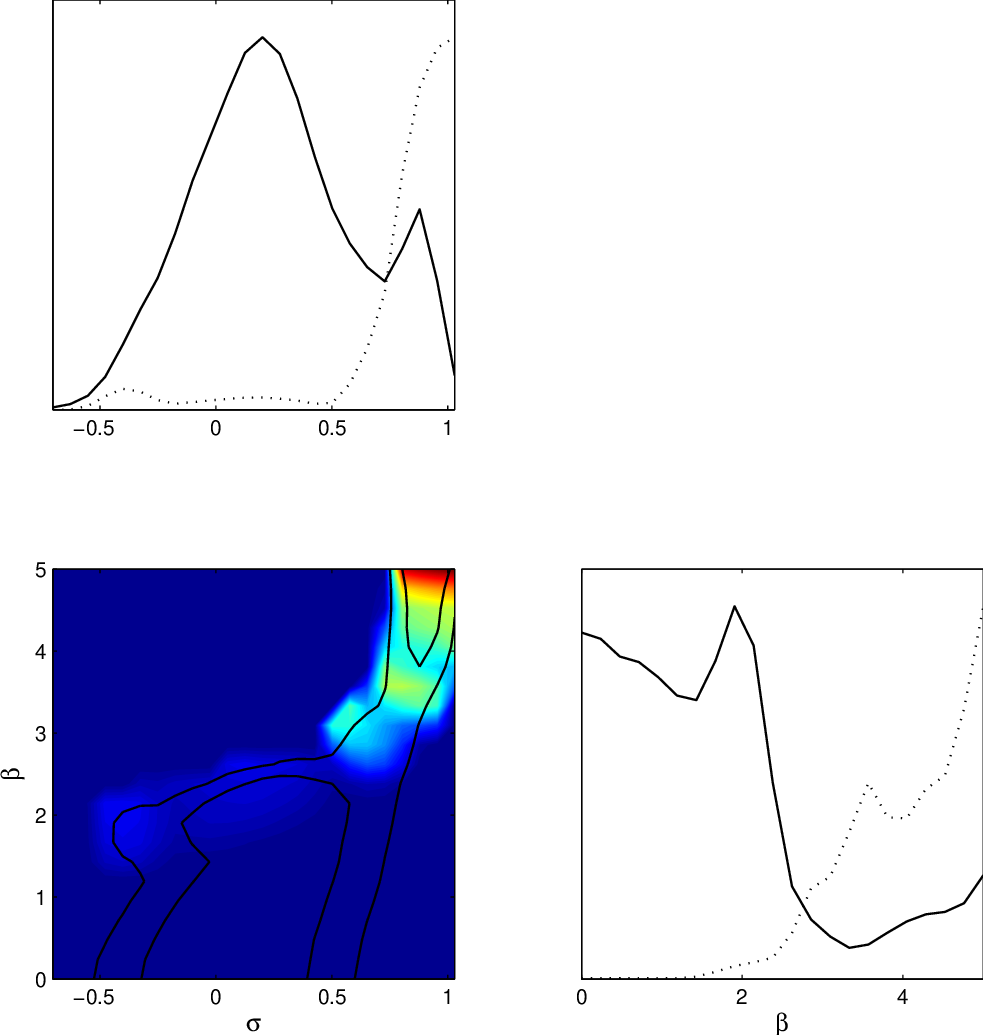}
\vskip.1truecm
\includegraphics[height=5.cm,angle=0]{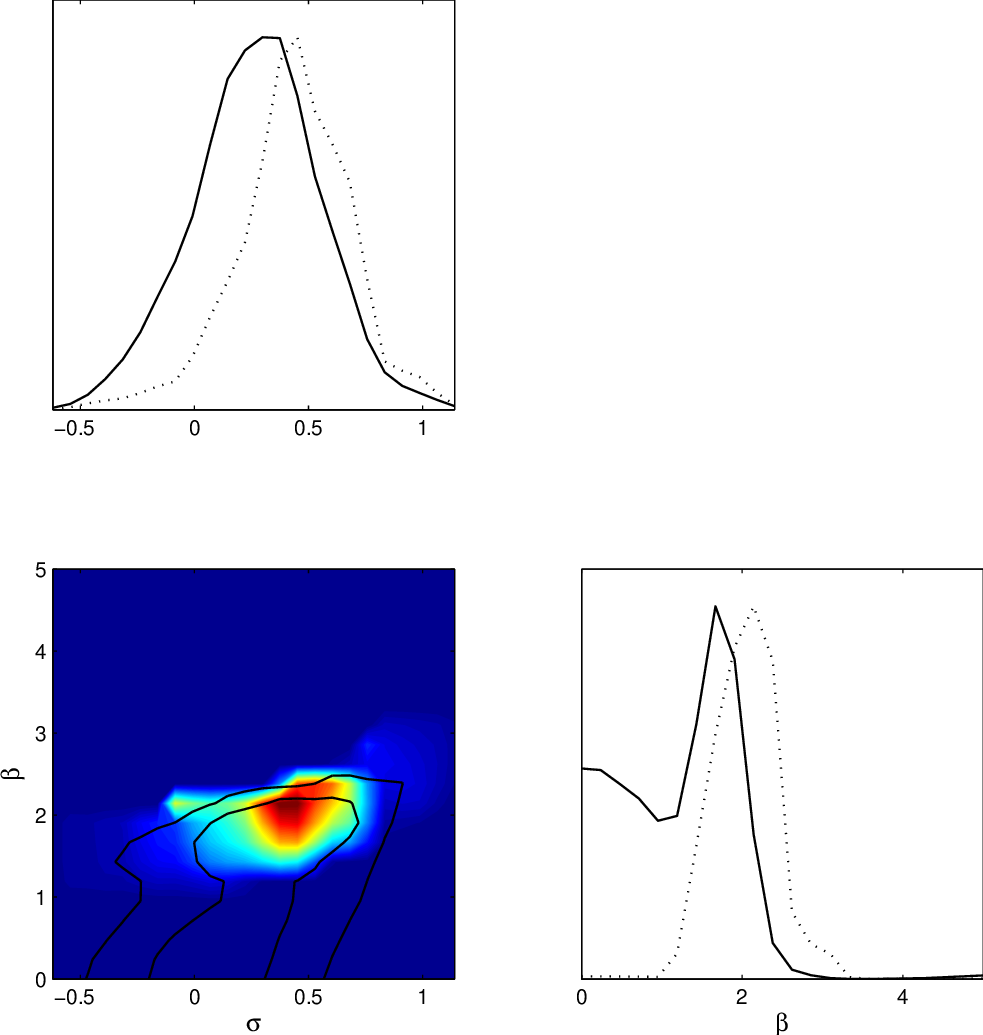}
\end{center}
\caption{Upper (lower) panel shows posterior constraints for the
exponential potential with growing $\nu$ mass. Here we fixed $m_\nu
= 0.2$eV ($m_\nu = 0.3$eV), $\omega_b = 0.022$, H$_0 = 72$}
\label{mav24}
\end{figure}

\begin{figure}%[!h]
\begin{center}
\includegraphics[height=5.cm,angle=0]{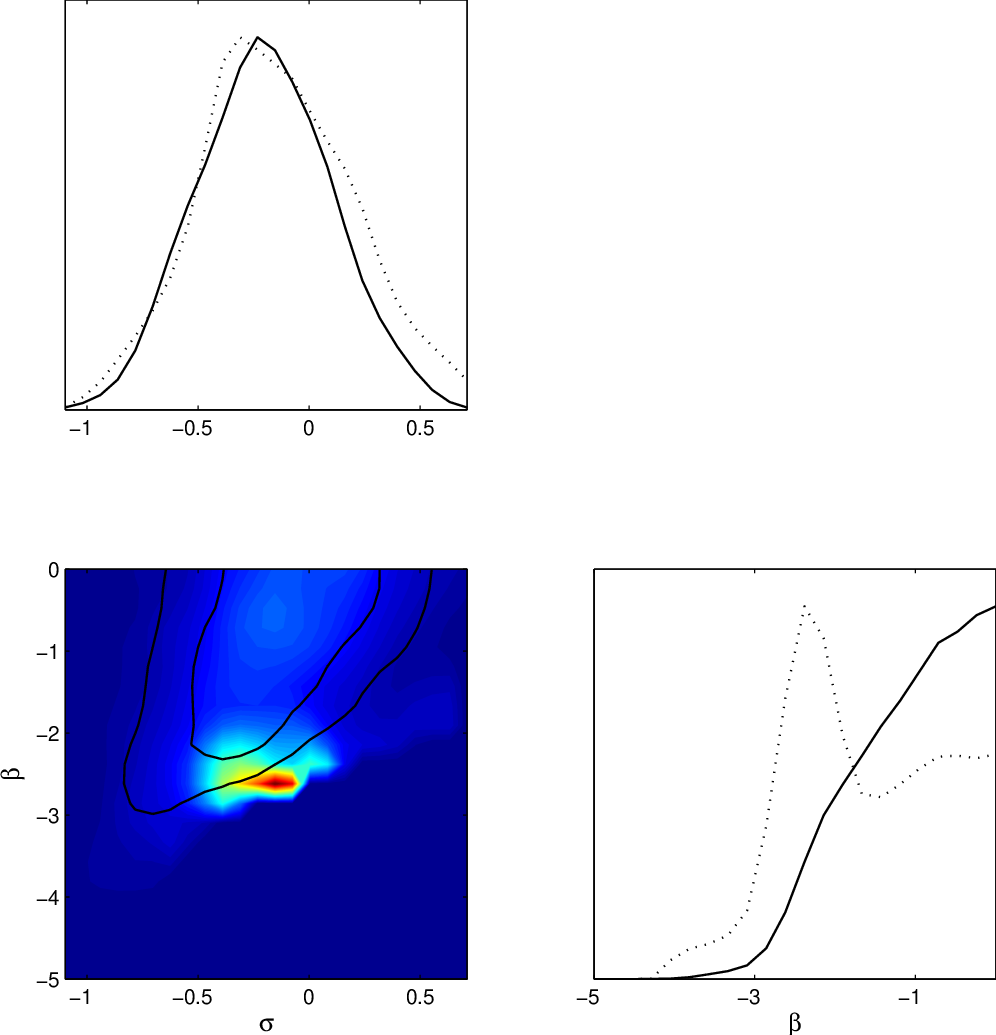}
\vskip.1truecm
\includegraphics[height=5.cm,angle=0]{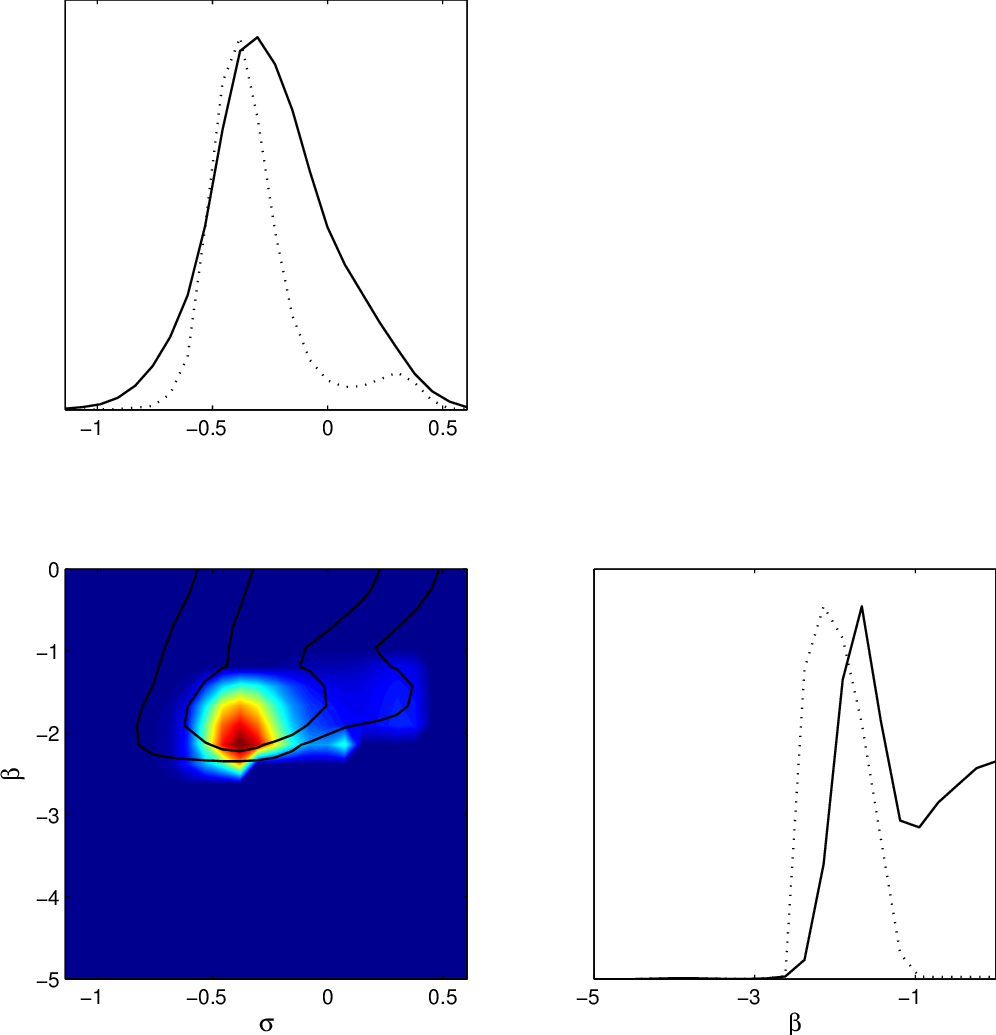}
\end{center}
\caption{Upper (lower) panel shows posterior constraints for the
exponential potential with decreasing $\nu$ mass. Here we fixed $m_\nu
= 0.2$eV ($m_\nu = 0.3$eV), $\omega_b = 0.022$, H$_0 = 72$}
\label{mav21}
\end{figure}

%In Fig.~\ref{mav24} and \ref{mav21} we show likelihood contours and
%distributions for growing and decreasing neutrino mass, respectively.
%The most significant result in both cases is the improvement in
%parameter constraints. In fact 
As shown in Fig.~\ref{mav24} and \ref{mav21}, the most significant
result is the improvement in the limits on the coupling $|\beta|$,
%become more stringent,
which has an upper value of the order between 2 or 3, and the distributions
for $\sigma$ become narrower, but are always compatible with zero at 68\%
CL.
%
%It is worth to notice that the coloured plot in the upper panel in
%Fig.~\ref{mav24} shows a complex structure for the likelihood contours
%in which the upper value of the coupling still exceeds $\beta=5$. This
%effect
The complex structure for the likelihood contours in the upper panel
of Fig.~\ref{mav24} can be mainly ascribed to the peculiar features of
the likelihood in Fig.~\ref{mav01}, which are completely lost when
considering a prior with higher mass.

%%%%%%%%%%%%%%%%%%%%%%%%%%%%%%%%%%%%%%%%%%%%%%%%%%%%%%%%%%%%%%%%%%%%
%%%%%%%%%%%%%%%%%%%%%%%%%%%%%%%%%%%%%%%%%%%%%%%%%%%%%%%%%%%%%%%%%%%%

%%%%%%%%%%%%%%%%%%%%%%%%%%%%%%%%%%%%%%%%%%%%%%%%%%%%%%%%%%%%%%%%%%%%
%%%%%%%%%%%%%%%%%%%%%%%%%%%%%%%%%%%%%%%%%%%%%%%%%%%%%%%%%%%%%%%%%%%%

\section{Fisher matrix forecasts}

%In the next few years new and more accurate datasets will be
%available. 
In this section, we show Fisher matrix results from the combination of
the CMB anisotropies and the tomographic weak lensing (TWL) spectra.
We considered the exponential potential with an exponential
coupling. The fiducial parameters $\theta_\alpha$ are shown in
Tab.~\ref{tab1}. These two sets correspond to the maximum area of the
likelihood as determined above and are chosen to be
exactly the same, except for the sign of $\beta$.

%%%%%%%%%%%%%%%%%%%%%%%%%%%%%%%%%%%%%%%%%%%%%%%%%%%%%%%%%%%%%%%%%%%%
\begin{table}%[!hbt]
\begin{center}
\caption{Fiducial cosmological parameters for the exponential
potential, which are consistent with Fig.~\ref{mav01}.} \label{tab1}
\begin{tabular}{ccccccccc}
\hline \hline Parameter&$\omega_b$& $\omega_{CDM}$& $H_0$& $z_{re}$&
${\rm n_s}$& $\ln{\rm A_s}$\\ Value&0.0224 & 0.1097 & 69 & 11 & 0.95 &
2.09$\times 10^{-9}$\\ \hline Parameter&$\omega_\nu$&$\sigma$& $\beta$
\\ Value&0.0010& 0.02& $\pm 1.9$ \\ \hline
\end{tabular}
\end{center}
\end{table}

The Fisher matrix for CMB power spectrum
\citep{Zaldarriaga:1996xe,Zaldarriaga:1997ch,Rassat:2008ja} is
calculated using Planck mission \citep{Planck:2006aa}
specifications.
%Note that Eq.~(\ref{eqn:cmbfisher}) usually includes a summation over
%the Planck frequency channels. However we conservatively 
We assume that we only use the 143 GHz channel as the science
channel. %, with the other frequencies used for foreground removal (not
%treated in this paper). 
This channel has a beam of $\theta_{\rm fwhm}=7.1'$ and sensitivities
of $\sigma_T = 2.2 \mu K/K$ and $\sigma_P = 4.2\mu K/K$. To account
for the galactic plane cut, we take $f_{\rm sky} = 0.80$. Note that we use
$\ell_{\rm min}=30$ as a minimum $\ell$-mode to avoid
problems with polarization foregrounds and subtleties when
modelling of the integrated Sachs-Wolfe effect.

%The Fisher matrix for TWL reads:
%% 
%\bea \lefteqn{F_{\alpha\beta}^{\rm WL} =}\label{derfish} \\ & &
%f_{sky}\sum_{\ell}^{\ell_{max}} \frac{(2\ell+1) \Delta\ell}{2}
%{\partial P_{ij,\ell} \over \partial \theta_{\alpha}} P_{jk,\ell}^{-1}
%{\partial P_{km,\ell} \over \partial \theta_{\beta}}\nonumber
%P_{mi,\ell}^{-1}~.  \eea
%% 
%where $P_{ij,\ell}$ are the components of the non-linear weak lensing
%power spectrum for the $i$-th and $j$-th bin \citep{Hu:2003pt}. In the
%expression~(\ref{derfish}) a summation over repeated indices is
%implicit. The redshift bins have been chosen such that each contains
%the same amount of galaxies. We consider the 5 bin case as reference
%case.

%The non-linear corrections are calculated using {\sc halofit}
%\citep{Smith:2002dz}. This procedure is suitably fitted to
%$\Lambda$CDM N-body simulations. Therefore it could lead to errors of
%the order of 20\% on Fisher outputs, if used for models different from
%$\Lambda$CDM (see {\it e.g.} \cite{Casarini:2011ms}). However, in the
%absence of suitable extensions for MaVaN, we assume {\sc halofit} as
%the procedure for non-linear corrections, reporting results up to
%$\ell_{max} = 1000$, in the mildly non-linear regime. This
%conservative choice also prevent us from considering a regime in which
%baryons could strongly affect matter power spectra and, as a
%consequence, weak lensing spectra \citep{Casarini:2012qj}.

The Fisher matrix for TWL \citep{Hu:2003pt} is calculated using the
forthcoming Euclid mission\footnote{http://www.euclid-ec.org}
specifications \citep{Laureijs:2011mu}. The survey area covered by the
experiment is $15000$~deg$^2$, while the density is 30 galaxies per
arcmin$^2$. The distribution of the galaxy number on the redshift and
solid angle is $n(z)~=~n_0z^2e^{-(z/z_0)^{1.5}}$ with a median
redshift $\bar z = 0.9$. The photometric redshift error is
$0.05~(1+z)$. We consider the five bin case as reference case.
Non-linear corrections of the matter power spectra are calculated using
{\sc halofit} \citep{Smith:2002dz}. Since this procedure is suitably
fitted to $\Lambda$CDM N-body simulations, it can lead to errors of
the order of 20\% on Fisher outputs, if used for models different from
$\Lambda$CDM (see {\it e.g.}  \cite{Casarini:2011ms}). However, we use {\sc halofit} in the
absence of suitable extensions for MaVaN, and
report results up to $\ell_{max} = 1000$ in the mildly non-linear
regime. This conservative choice also prevents us from the effects of
baryons on the matter power spectra and weak lensing spectra
\citep{Casarini:2012qj}.

%%%%%%%%%%%%%%%%%%%%%%%%%%%%%%%%%%%%%%%%%%%%%%%%%%%%%%%%%%%%%%%%%%%%
\begin{table}%[!hbt]
\begin{center}
\caption{1-$\sigma$ error estimations of cosmological parameters for
  an exponential potential with $\beta<0$, using Planck-like CMB and
  Euclid-like Tomographic Weak Lensing (TWL) data.\label{tab2}}
\begin{tabular}{ccccccccc}
\hline
\hline
Dataset&$\sigma_{\omega_\nu}$&$\sigma_{\sigma}$& $\sigma_{\beta}$ \\
\hline
TWL&0.00043& 0.024& 1.17 \\
CMB&0.00064& 0.014& 1.28 \\
TWL+CMB&0.00025& 0.012& 0.69 \\
\hline
\end{tabular}
\end{center}
\end{table}
%%%%%%%%%%%%%%%%%%%%%%%%%%%%%%%%%%%%%%%%%%%%%%%%%%%%%%%%%%%%%%%%%%%%

\begin{table}%[!hbt]
\begin{center}
\caption{The same as Table~\ref{tab2}, but with $\beta>0$. \label{tab3}}
\begin{tabular}{ccccccccc}
\hline
\hline
Dataset&$\sigma_{\omega_\nu}$&$\sigma_{\sigma}$& $\sigma_{\beta}$ \\
\hline
TWL&0.00045& 0.028& 2.04 \\
CMB&0.00054& 0.019& 1.71 \\
TWL+CMB&0.00026& 0.012& 1.15 \\
\hline
\end{tabular}
\end{center}
\end{table}
%%%%%%%%%%%%%%%%%%%%%%%%%%%%%%%%%%%%%%%%%%%%%%%%%%%%%%%%%%%%%%%%%%%%

%%%%%%%%%%%%%%%%%%%%%%%%%%%%%%%%%%%%%%%%%%%%%%%%%%%%%%%%%%%%%%%%%%%%
\begin{figure}%[!h]
\begin{center}
\includegraphics[height=5.cm,angle=0]{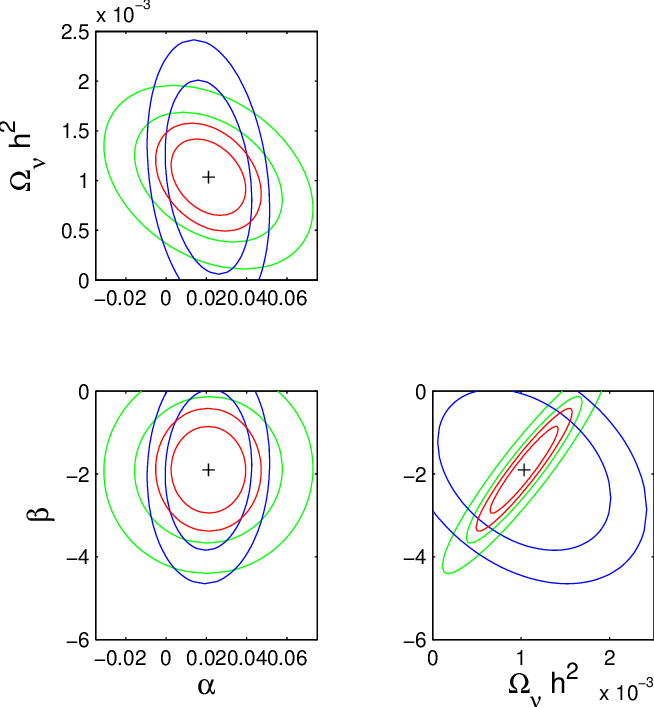}
\vskip.1truecm
\includegraphics[height=5.cm,angle=0]{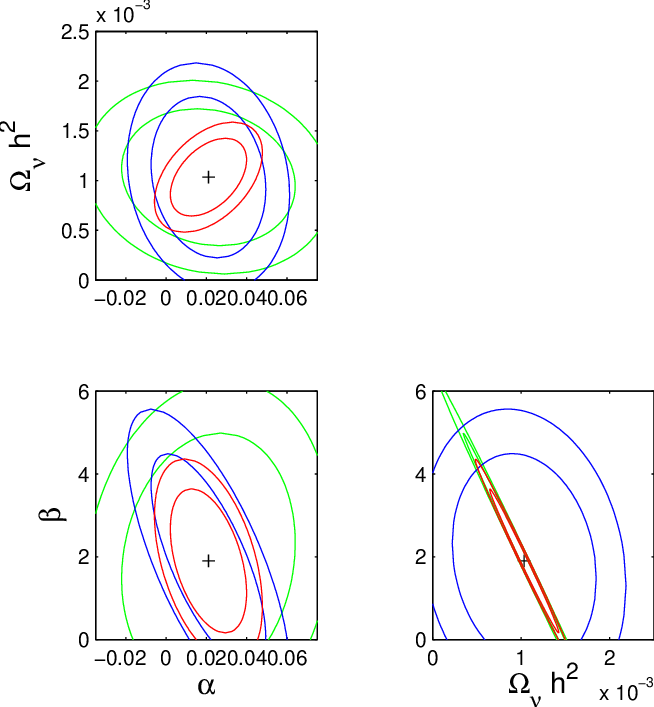}
\end{center}
\caption{Upper (lower) panel shows 
%likelihood contour forecast at
68\% and 95\% likelihood confidence levels for the exponential
potential, in the decreasing (growing) mass case. The green lines are
obtained considering an Euclid-like TWL survey.
%, with 5 bins and $\ell_{max}=1000$.
The blue lines are for a Planck-like CMB mission. The red lines are
for the combination of the two.}
\label{fisher1}
\end{figure}
%%%%%%%%%%%%%%%%%%%%%%%%%%%%%%%%%%%%%%%%%%%%%%%%%%%%%%%%%%%%%%%%%%%%

%%%%%%%%%%%%%%%%%%%%%%%%%%%%%%%%%%%%%%%%%%%%%%%%%%%%%%%%%%%%%%%%%%%%
\begin{figure}%[!h]
\begin{center}
\includegraphics[height=5.cm,angle=0]{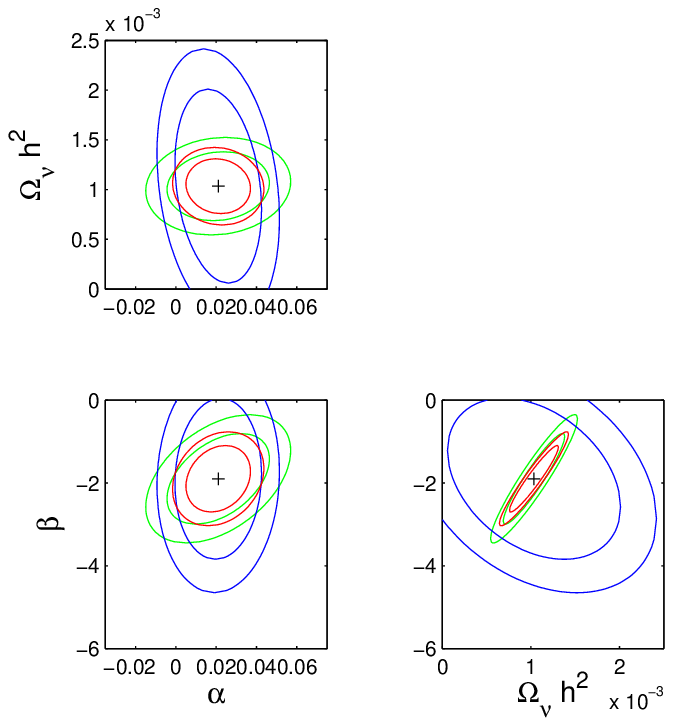}
\vskip.1truecm
\includegraphics[height=5.cm,angle=0]{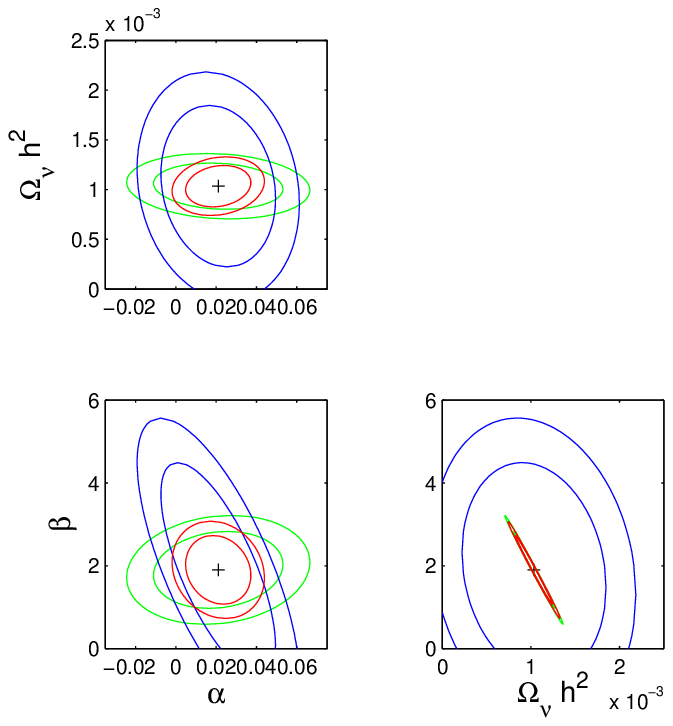}
\end{center}
\caption{As Fig~\ref{fisher1} with 5 bins and $\ell_{max}=5000$.}
\label{fisher2}
\end{figure}
%%%%%%%%%%%%%%%%%%%%%%%%%%%%%%%%%%%%%%%%%%%%%%%%%%%%%%%%%%%%%%%%%%%%

%The error estimation of DE and neutrino density parameters is reported
%for a MaVaN model with an exponential coupling for growing and
%decreasing neutrino mass. 
As shown in Table~\ref{tab2}, TWL is able to put constraints stronger
than CMB on $\beta<0$ and $\omega_\nu$, with an improvement of
$\sim$10\% and $>$30\%, respectively. Instead CMB is more
efficient in constraining $\sigma$, gaining $\sim$50\% over TWL. The
combination of the two observables can improve constraints on all
parameters.
%In this case, the main result is for
%the estimate of $\omega_\nu$ constraint to $0.00025$,
%$\sigma_{m_\nu}\sim 0.02$eV. 
Analogous comments can be made for the $\beta>0$ in Table~\ref{tab3}
with the only difference that CMB can constrain $\beta$ better than
TWL.

In Fig.~\ref{fisher1}, we show 1-$\sigma$ and 2-$\sigma$ likelihood
contours %for MaVan's model with an exponential coupling for both
%growing and decreasing neutrino mass. Here we report results 
for $\omega_\nu$, $\beta$ and $\sigma$, after marginalising over the
other parameters. It is clear that neither Planck nor Euclid
alone can be able to constrain a non-zero coupling $\beta\sim{\cal
  O}(1)$; only TWL is slightly more efficient than CMB in the
$\beta<0$ case. In this case the combination of the two can exclude a
null coupling with a significance higher than 2-$\sigma$.
We note that 
%we performed a conservative analysis of TWL Fisher
%matrix, using multipoles up to $\ell_{max} = 1000$. 
%The capability of
%TWL in constraining $\beta$ and the other parameters will be
%significantly increased by the use of higher multipoles, delving into
%the deep non-linear regime. 
higher multipoles can significantly reduce the limits on $\beta$ and
the other parameters. If we suppose, for instance, that {\sc halofit} is
appropriate for our models at multipoles up to $\ell_{max} = 5000$ in
the TWL Fisher matrix (see Fig.~\ref{fisher2}), the estimated errors are then
clearly reduced with a factor $\sim$2 with respect to
Fig.~\ref{fisher2}.
% for Weak Lensing and, as
%a consequence, for the combined case. 
In this case, Euclid would be able to exclude the zero value both for
$\beta$ and $\omega_\nu$ with high statistical significance. 
%However
%in this case an appropriate extension of {\sc halofit} would be
%necessary, including both appropriate non-linear correction for MaVaN
%models and baryon physics.

Despite this limitation, a very interesting result is that
TWL, whether alone or in combination with CMB, is able to find a lower value
of $\Omega_\nu h^2$ with an error that is half of the CMB error if taken
alone. Moreover, a strong degeneracy can be found in the
$\omega_\nu$-$\beta$ plane with regard to the TWL ellipses and also in
combination with CMB. This effect appears in both cases in a symmetric
way, showing that lower $|\beta|$ values allow higher neutrino
mass. The $\Omega_\nu h^2$-$\beta$ degeneracy could be eventually
broken from external priors on $m_\nu$. On the other hand, no
noteworthy remark can be said about the $\sigma$ parameter, whose
constrain remains compatible with zero for the combination of
the two observables.

\section{Discussion and conclusions}

In this work, we focused on the hypothesis that the origin of
cosmic acceleration can be attributed to a quintessence scalar field
coupled to massive neutrinos. First, we updated and extended
parameter constraints using the most recent available data from SNIa,
CMB, and LSS observations. We considered both an exponential and a
power law potential with growing ($\beta>0$) or decreasing
($\beta<0$) neutrino mass.

The cosmological data did not place strong constraints
on the coupling parameter in the low coupling range or for the
neutrino density parameter, on which only upper limits can be
placed in any of these cases. The main outcome of the analysis was that $\beta$ values $\sim
{\cal O}(1)$ are compatible with actual data with a neutrino
mass $m_\nu\lesssim 0.32$eV.

Therefore we do have not enough information at the moment to exclude a
possible coupling between the neutrino and DE field in either the low or
the high coupling regime. New and precise data from observables
related to the recent evolution of the universe are necessary
with a deeper understanding of the MaVaN theory dynamics. In this
sense, the forthcoming Euclid weak lensing and galaxy clustering data
represent the turning point for the future progress in cosmology.

With this aim, a Fisher matrix study was accomplished by considering
future data release from missions, such as Planck for CMB spectra and
Euclid for tomographic weak lensing spectra. The latter experiment
is more efficient in improving present constraints on
$\omega_\nu$, even considering a cautious use of multipoles up to
$\ell_{max} = 1000$. This choice prevents us from including highly
non-linear features, which are not correctly predicted by {\sc
halofit} for MaVaN theories. Combining Euclid data with complementary
information from Planck reduces the estimated error to about a
factor 2 with respect to considering the two datasets alone. It is
worth to mention that the strong degeneracy between $\beta$ and
$\omega_\nu$ could eventually be broken by an external prior on
the neutrino mass.

A crucial point for the forthcoming Euclid mission is to study
and implement an efficient way of including non-linear
corrections on the matter power spectrum calculations with {\sc halofit}
being only suitable for $\Lambda$CDM cosmologies. Delving
into the deep non-linear regime of matter perturbations could
significantly improve parameter estimation by fully exploiting the high
potentiality of the TWL from Euclid.

\begin{acknowledgements}
%GLV and DFM thank Valeria Pettorino for useful comments and
%discussions.
DFM thanks the Research Council of Norway. 
%FRINAT grant
%197251/V30. DFM is also partially supported by project
%CERN/FP/123615/2011 and PTDC/FIS/111725/2009.
\end{acknowledgements}

\bibliographystyle{aa}
\bibliography{mavbib}

\begin{thebibliography}{34}
\expandafter\ifx\csname natexlab\endcsname\relax\def\natexlab#1{#1}\fi

\bibitem[{Amanullah {et~al.}(2010)Amanullah, Lidman, Rubin, Aldering, Astier,
  {et~al.}}]{Amanullah:2010vv}
Amanullah, R., Lidman, C., Rubin, D., {et~al.} 2010, Astrophys.J., 716, 712

\bibitem[{Amendola {et~al.}(2008{\natexlab{a}})Amendola, Baldi, \&
  Wetterich}]{Amendola_Baldi_Wetterich_2008}
Amendola, L., Baldi, M., \& Wetterich, C. 2008{\natexlab{a}}, Phys.Rev., D78,
  023015

\bibitem[{Amendola {et~al.}(2008{\natexlab{b}})Amendola, Baldi, \&
  Wetterich}]{Amendola07}
Amendola, L., Baldi, M., \& Wetterich, C. 2008{\natexlab{b}}, Phys.Rev., D78,
  023015

\bibitem[{Binetruy(1999)}]{Binetruy:1998rz}
Binetruy, P. 1999, Phys.Rev., D60, 063502

\bibitem[{Bjaelde {et~al.}(2008)Bjaelde, Brookfield, van~de Bruck, Hannestad,
  Mota, {et~al.}}]{Brookfield1}
Bjaelde, O.~E., Brookfield, A.~W., van~de Bruck, C., {et~al.} 2008, JCAP, 0801,
  026

\bibitem[{Brookfield {et~al.}(2006{\natexlab{a}})Brookfield, van~de Bruck,
  Mota, \& Tocchini-Valentini}]{Brookfield3}
Brookfield, A., van~de Bruck, C., Mota, D., \& Tocchini-Valentini, D.
  2006{\natexlab{a}}, Phys.Rev.Lett., 96, 061301

\bibitem[{Brookfield {et~al.}(2006{\natexlab{b}})Brookfield, van~de Bruck,
  Mota, \& Tocchini-Valentini}]{Brookfield2}
Brookfield, A.~W., van~de Bruck, C., Mota, D., \& Tocchini-Valentini, D.
  2006{\natexlab{b}}, Phys.Rev., D73, 083515

\bibitem[{Casarini {et~al.}(2012)Casarini, Bonometto, Borgani, Dolag, Murante,
  {et~al.}}]{Casarini:2012qj}
Casarini, L., Bonometto, S.~A., Borgani, S., {et~al.} 2012, arXiv:1203.5251
  [astro-ph.CO]

\bibitem[{Casarini {et~al.}(2011)Casarini, La~Vacca, Amendola, Bonometto, \&
  Maccio}]{Casarini:2011ms}
Casarini, L., La~Vacca, G., Amendola, L., Bonometto, S.~A., \& Maccio, A.~V.
  2011, JCAP, 1103, 026

\bibitem[{Dickinson {et~al.}(2004)Dickinson, Battye, Carreira, Cleary, Davies,
  {et~al.}}]{Dickinson:2004yr}
Dickinson, C., Battye, R.~A., Carreira, P., {et~al.} 2004,
  Mon.Not.Roy.Astron.Soc., 353, 732

\bibitem[{Fisher(1935)}]{Fisher:1935}
Fisher, R.~A. 1935, J. Roy. Statist. Soc., 98, 39

\bibitem[{Hu \& Jain(2004)}]{Hu:2003pt}
Hu, W. \& Jain, B. 2004, Phys.Rev., D70, 043009

\bibitem[{Klapdor-Kleingrothaus {et~al.}(2004)Klapdor-Kleingrothaus,
  Krivosheina, Dietz, \& Chkvorets}]{KlapdorKleingrothaus:2004wj}
Klapdor-Kleingrothaus, H., Krivosheina, I., Dietz, A., \& Chkvorets, O. 2004,
  Phys.Lett., B586, 198

\bibitem[{Klapdor-Kleingrothaus(2005)}]{KlapdorKleingrothaus:2005qv}
Klapdor-Kleingrothaus, H.~V. 2005, 215, hep-ph/0512263

\bibitem[{Komatsu {et~al.}(2011)}]{Komatsu:2010fb}
Komatsu, E. {et~al.} 2011, Astrophys.J.Suppl., 192, 18

\bibitem[{Laureijs {et~al.}(2011)Laureijs, Amiaux, Arduini, Augueres,
  Brinchmann, {et~al.}}]{Laureijs:2011mu}
Laureijs, R., Amiaux, J., Arduini, S., {et~al.} 2011, arXiv:1110.3193
  [astro-ph.CO]

\bibitem[{Lewis \& Bridle(2002)}]{Lewis:2002ah}
Lewis, A. \& Bridle, S. 2002, Phys. Rev., D66, 103511

\bibitem[{Lewis {et~al.}(2000)Lewis, Challinor, \& Lasenby}]{camb}
Lewis, A., Challinor, A., \& Lasenby, A. 2000, Astrophys.J., 538, 473

\bibitem[{Mota {et~al.}(2008)Mota, Pettorino, Robbers, \& Wetterich}]{grow2}
Mota, D., Pettorino, V., Robbers, G., \& Wetterich, C. 2008, Phys.Lett., B663,
  160

\bibitem[{Pettorino {et~al.}(2009)Pettorino, Mota, Robbers, \&
  Wetterich}]{grow3}
Pettorino, V., Mota, D.~F., Robbers, G., \& Wetterich, C. 2009, AIP Conf.Proc.,
  1115, 291

\bibitem[{Pettorino {et~al.}(2010)Pettorino, Wintergerst, Amendola, \&
  Wetterich}]{Pettorino:2010bv}
Pettorino, V., Wintergerst, N., Amendola, L., \& Wetterich, C. 2010, Phys.Rev.,
  D82, 123001

\bibitem[{Rassat {et~al.}(2008)Rassat, Amara, Amendola, Castander, Kitching,
  {et~al.}}]{Rassat:2008ja}
Rassat, A., Amara, A., Amendola, L., {et~al.} 2008, arXiv:0810.0003 [astro-ph]

\bibitem[{Reichardt {et~al.}(2009)Reichardt, Ade, Bock, Bond, Brevik,
  {et~al.}}]{Reichardt:2008ay}
Reichardt, C., Ade, P., Bock, J., {et~al.} 2009, Astrophys.J., 694, 1200

\bibitem[{Riess {et~al.}(2009)Riess, Macri, Casertano, Sosey, Lampeitl,
  {et~al.}}]{Riess:2009pu}
Riess, A.~G., Macri, L., Casertano, S., {et~al.} 2009, Astrophys.J., 699, 539

\bibitem[{Sievers {et~al.}(2007)Sievers, Achermann, Bond, Bronfman, Bustos,
  {et~al.}}]{Sievers:2005gj}
Sievers, J.~L., Achermann, C., Bond, J., {et~al.} 2007, Astrophys.J., 660, 976

\bibitem[{Smith {et~al.}(2003)}]{Smith:2002dz}
Smith, R. {et~al.} 2003, Mon.Not.Roy.Astron.Soc., 341, 1311

\bibitem[{Sturm(2011)}]{Sturm:2011ms}
Sturm, M. 2011, arXiv:1111.4773 [hep-ex]

\bibitem[{Tegmark {et~al.}(1997)Tegmark, Taylor, \& Heavens}]{Tegmark:1996bz}
Tegmark, M., Taylor, A., \& Heavens, A. 1997, Astrophys.J., 480, 22

\bibitem[{Tegmark {et~al.}(2006)}]{Tegmark:2006az}
Tegmark, M. {et~al.} 2006, Phys.Rev., D74, 123507

\bibitem[{{The Planck Collaboration}(2006)}]{Planck:2006aa}
{The Planck Collaboration}. 2006, arXiv:astro-ph/0604069

\bibitem[{Wetterich(1988)}]{Wetterich_1988}
Wetterich, C. 1988, Nuclear Physics B, 302, 645

\bibitem[{Wintergerst {et~al.}(2010)Wintergerst, Pettorino, Mota, \&
  Wetterich}]{grow5}
Wintergerst, N., Pettorino, V., Mota, D., \& Wetterich, C. 2010, Phys.Rev.,
  D81, 063525

\bibitem[{Zaldarriaga \& Seljak(1997)}]{Zaldarriaga:1996xe}
Zaldarriaga, M. \& Seljak, U. 1997, Phys.Rev., D55, 1830

\bibitem[{Zaldarriaga {et~al.}(1997)Zaldarriaga, Spergel, \&
  Seljak}]{Zaldarriaga:1997ch}
Zaldarriaga, M., Spergel, D.~N., \& Seljak, U. 1997, Astrophys.J., 488, 1

\end{thebibliography}

\end{document}